\documentclass[12pt,preprint]{aastex}
\newcommand{\mincir}{\raise
-3.truept\hbox{\rlap{\hbox{$\sim$}}\raise4.truept\hbox{$<$}\ }}
\newcommand{\magcir}{\raise
-3.truept\hbox{\rlap{\hbox{$\sim$}}\raise4.truept\hbox{$>$}\ }}
\newcommand{\minmag}{\raise
-3.truept\hbox{\rlap{\hbox{$<$}}\raise5.truept\hbox{$<$}\ }}
\newcommand{\be}{\begin{equation}}
\newcommand{\ee}{\end{equation}}
\newcommand{\ba}{\begin{eqnarray}}
\newcommand{\ea}{\end{eqnarray}}
\newcommand{\brr}{\begin{array}}
\newcommand{\err}{\end{array}}
\newcommand{\bc}{\begin{center}}
\newcommand{\ec}{\end{center}}

\newenvironment{inlinefigure}{%
\def\@captype{inlinefigure}%
\noindent\begin{minipage}{\linewidth}\begin{center}}
{\end{center}\end{minipage}\smallskip}
\makeatother

\shorttitle{Clustering of hard X-ray selected sources} 
\shortauthors{S. Basilakos et al.}

\begin{document}

\title{The Clustering of XMM-{\em Newton} Hard X-ray Sources}

\author{S. Basilakos\altaffilmark{1}, A. Georgakakis\altaffilmark{1}, 
M. Plionis\altaffilmark{1,2}, I. Georgantopoulos\altaffilmark{1}}
\altaffiltext{1}{Institute of Astronomy \& Astrophysics, 
National Observatory of Athens, I.Metaxa \& B.Pavlou, 
P.Penteli 152 36, Athens, Greece; svasil@astro.noa.gr, 
age@astro.noa.gr, ig@astro.noa.gr}
\altaffiltext{2}{Instituto Nacional de Astrofisica, 
Optica y Electronica (INAOE)
Apartado Postal 51 y 216, 72000, Puebla, Pue., Mexico; mplionis@inaoep.mx}

\begin{abstract}
This paper presents the clustering properties of hard (2-8\,keV) X-ray
selected sources detected in a wide field ($\approx \rm 2\,deg^{2}$)
shallow [$f_X(\rm 2-8\,keV)\approx 10^{-14}\rm \, erg \, cm^{-2} \,
s^{-1}$] and contiguous  XMM-{\it Newton} survey. We perform an angular
correlation function analysis using a total of  171 sources to the
above flux limit.  We detect a $\sim 4\sigma$ correlation signal out to
300\,arcsec with $w(\theta < 300^{''})\simeq  0.13 \pm 0.03$. Modeling
the two point correlation function as a  power law of the form 
$w(\theta)=(\theta_{\circ}/\theta)^{\gamma-1}$ we find:
$\theta_{\circ}=48.9^{+15.8}_{-24.5}$ arcsec and $\gamma=2.2\pm
{0.30}$. Fixing the correlation function slope to $\gamma=1.8$ we
obtain  $\theta_{\circ}=22.2^{+9.4}_{-8.6}$\,arcsec. Using 
Limber's intergral equation and a variety of possible luminosity functions of
the hard X-ray population, we find a relatively large correlation
length, ranging from $r_{\circ}\sim 9$ to 19 $h^{-1}$ Mpc (for
$\gamma=1.8$ and the {\em concordance} cosmological model), with this range
reflecting also different evolutionary models for the source
luminosities and clustering characteristics.
%For example, in the concordance cosmological model and assuming
%comoving clustering and a
%luminosity dependent density evolution model
%for the luminosity function (Ueda et al 2003) we obtain a
%correlation length of $r_{\circ} \simeq 19 \pm 3 \; h^{-1}$ 
%and $13\pm 3 \; h^{-1}$\, Mpc  for $\gamma=1.8$.
% and  $\gamma=2.2$ respectively.
The relatively large correlation length is comparable
to that of extremely red objects and luminous radio sources.
\end{abstract}


\section{Introduction}
It is well known that the study of the distribution of matter on
large scales, using different extragalactic objects
provides important constraints on models of cosmic structure formation.
Since Active Galactic Nuclei (AGN) can be detected up 
to very high redshifts they provide information on the
underlying mass distribution as well as on the 
evolution of large scale structure  
(cf. Hartwick \& Schade 1989; Basilakos 2001 and references therein). 

The traditional indicator of clustering, the 
angular two-point correlation function, is a fundamental and simple 
statistical test
for the study of any extragalactic mass tracer and is relatively
straightforward to measure from observational data. 
The overall knowledge of the AGN clustering using X-ray data 
comes mostly from the soft ($\le 3$keV) X-ray band 
(Boyle \& Mo 1993; Vikhlinin \& Forman 1995; Carrera et al. 1998; 
Akylas, Georgantopoulos, Plionis, 2000; Mullis 2002), which is however biased
against absorbed AGNs. Hard X-ray surveys ($\ge 2$keV) play a
key role in our understanding of how the whole AGN population,
including obscured (type II) AGNs, trace the  
underlying mass distribution. Furthermore, understanding the 
spatial distribution of type II AGNs is important since
they are among the main contributors of the cosmic X-ray background 
(Mushotzky et al. 2000;
Hasinger et al. 2001; Giacconi et al. 2002).

Recently, Yang et al. (2003)   
performing a counts-in cells analysis of a
deep ($f_{2-8 keV} \sim 3 \times 10^{-15}$ erg s$^{-1}$ cm$^{-2}$)  
{\em Chandra} survey in the Lockman Hole North-West region,
found that the hard band sources are highly clustered 
with $\sim$ 60$\%$ of them being distributed in overdense regions.
The XMM-{\em Newton} with $\sim 5$ times more effective area,
especially at hard energies, and $\sim 3$ times larger field of view 
(FOV) provides an ideal instrument for clustering studies of X-ray sources.

In this paper we estimate for the first time the angular correlation
function of the XMM-{\em Newton} hard X-ray sample. Using Limber's
equation and different models of the luminosity funtion for these sources
we derive the expected spatial correlation function which we compare
with that of a variety of extragalactic populations.
Hereafter, all $H_{\circ}$-dependent quantities will be given in units of 
$h\equiv H_{\circ}/100$ km $s^{-1}$ Mpc$^{-1}$.

\section{The Sample}
The hard X-ray selected sample used in the present study is compiled
from the XMM-{\it Newton}/2dF survey. This is a shallow (2-10\,ksec per
pointing) survey carried out by the XMM-{\it Newton} near the North
Galactic Pole  [NGP; RA(J2000)=$13^{\rm h}41^{\rm m}$;  
Dec.(J2000)=$00\degr00\arcmin$] and the South Galactic Pole
[SGP; RA(J2000)=$00^{\rm h} 57^{\rm m}$, Dec.(J2000)=$-28\degr
00\arcmin$] regions. A total of 18 XMM-{\it Newton} pointings were
observed equally split between the NGP and the SGP areas. However, a
number of pointings were discarded due to elevated particle
background at the time of  the observation. This  results in a total
of 13 usable XMM-{\it Newton} pointings covering an area of $\rm
\approx2\,deg^{2}$. A full description of the data  reduction, source
detection and flux estimation are presented by Georgakakis et
al. (2003, 2004).  For the 2-D correlation analysis presented in this
paper we use the hard (2-8\,keV) band catalogue of the XMM-{\it
Newton}/2dF survey. This comprises a total of 171 sources above the
$5\sigma$  detection threshold to the limiting flux of $f_X(\rm
2-8\,keV) \approx 10^{-14}\, erg \, s^{-1}\, cm^{-2}$. Note that our
hard X-ray sources comprise of a mixture of QSO's and relatively
nearby ($z<0.8$) galaxies with red colours $g-r>0.5$ which are most probably
associated with obscured low-luminosity AGN (Georgantopoulos et al. 2004).

\section{Correlation function analysis}
\subsection{The angular correlation}
The clustering properties of the hard X-ray selected sources are
estimated using the two point angular correlation function $w(\theta)$
defined by $dP=n^{2}[1+w(\theta)]d\Omega_{1}d\Omega_{1}$,
where $dP$ is the joint probability of finding two sources in the
solid  angle elements $d\Omega_{1}$ and  $d\Omega_{2}$ separated by 
angle $\theta$ and $n$ is the mean surface density of sources. For a
random distribution of sources $w(\theta)$=0. Therefore, the  angular
correlation function provides a measure of galaxy density excess over
that expected for a random distribution. A variety of estimators
of $w(\theta)$ have been used over the years
(cf. Infante 1994). 

%In the present study, a source is taken as the `center' and
%the number of pairs within annular rings is  counted. 
In the present study we use the estimator (cf. Efstathiou et al. 1991):
\begin{equation}
w(\theta)=f(N_{DD}/N_{DR})-1,
\end{equation}
where $N_{DD}$ and $N_{DR}$ is the number of data-data and data-random
pairs respectively at separations $\theta$ and  $\theta +
d\theta$. In the above relation $f$ is the normalization factor
$f = 2 N_R /(N_D-1)$ where $N_D$ and $N_R$ are the total number of
data and random points respectively. The uncertainty in  
$w(\theta)$ is estimated as $\sigma_{w}=\sqrt{(1+w(\theta))/N_{DR}}$
(Peebles 1973). 
To account for the different source selection
and edge effects, we have produced 100 Monte Carlo random realizations of the 
source distribution within the area of the survey by taking into
account variations in sensitivity which might affect the correlation
function estimate. Indeed, the flux threshold for detection depends on
the off-axis angle from the center of each of the XMM-{\it Newton} 
pointings. Since
the random catalogues must have the same selection effects as the real 
catalogue, sensitivity maps are used to discard random points in
less sensitive areas (close to the edge of the pointings). 
This is accomplished, to  the first approximation, by
assigning a flux to the random points using the Baldi et al. (2002)
2-10\,keV $\log N - \log S$ (after transforming to the 2-8\,keV band
assuming $\Gamma=1.7$). If the  flux of a random point is less than
5 times the local {\em rms}  noise (assuming Poisson statistics for the
background) the point is excluded from the random data-set. We note
that the Baldi et al. (2002) $\log N - \log S$ is in good agreement
with the 2-8\,keV number counts estimated in the present 
survey. This is demonstrated in Figure 1 were we plot
our differential number counts and the best fit relation
of Baldi et al. (2002). Note that we have tested that
our random simulations reproduce 
both the off-axis sensitivity of the detector as well as 
the individual field $\log N - \log S$.

We apply the correlation analysis evaluating $w(\theta)$ in 
logarithmic intervals with $\delta \log \theta= 0.05$. The results are
shown in Figure 2, were the line corresponds to the best-fit power law model 
$w(\theta)=(\theta_{\circ}/\theta)^{\gamma-1}$ using the standard
$\chi^{2}$ minimization procedure in which each correlation point is
weighted by its error. 
We find a statistically significant signal 
with $w(\theta < 300^{''})\simeq  0.13 \pm 0.03$
at the 4.3$\sigma$ and $\sim 2.7 \sigma$ 
confidence level using Poissonian or bootstrap errors respectively.
Note that the bootstrap
errors probably overestimate the true uncertainty, especially in
sparse samples (Fisher et al 1994). Therefore 
the true significance level is somewhere in between the above two
values.

In the insert of Figure 2 we present the iso-$\Delta \chi^{2}$ contours 
(where $\Delta \chi^{2}=\chi^{2}-\chi_{\rm min}^{2}$) in the 
$\gamma-\theta_{\circ}$ plane. The contours correspond to
$1\sigma$ ($\Delta \chi^{2}=2.30$) and $2\sigma$ ($\Delta
\chi^{2}=6.17$) uncertainties, respectively.  
The best fit clustering parameters are:
\be\label{eq:res1}
\theta_{\circ}=48.9^{+15.8}_{-24.5} \;\; {\rm arcsec} \;\;\; \gamma=2.2\pm
{0.30}
\ee
where the errors correspond to $1\sigma$ ($\Delta \chi^{2}=2.30$) 
uncertainties. Fixing the correlation function slope to its
nominal value, $\gamma=1.8$, we estimate
$\theta_{\circ}=22.2^{+9.4}_{-8.6}$\,arcsec
\footnote{The robustness of our results to  
the fitting procedure was tested using different bins (spanning from
10 to 20) and no significant difference was found.}.
Note that our results do not suffer from the {\em
  amplification bias} which results from merging close
source pairs when the PSF size is larger than their typical 
separation (see Vikhlinin \& Forman 1995). This is because the 
estimated $\theta_{\circ}$ values are much larger than the XMM-{\em Newton}
PSF size of 6$^{''}$ FWHM.

Another systematic effect that could bias the 
angular 2-point correlation function is that introduced by the
so called {\em integral constraint}. This results from the fact that
the correlation function is estimated from a limited area, which
in turn implies that over the area studied the relation 
$\int \int w(\theta_{12})
{\rm d}\Omega_1 {\rm d}\Omega_2 =0$ should be satisfied. We can 
attempt to estimate the 
resulting underestimation
of the true correlation function by calculating the quantity: 
$W= \int {\rm d}\Omega_1 \int {\rm d}\Omega_2 w(\theta_{12}) / 
\int {\rm d}\Omega_1 \int {\rm d}\Omega_2$
Clearly, evaluating $W$ necessitates {\em a
  priori} knowledge of the angular correlation function. A 
tentative value of $W$ using a range of $w(\theta)$ given by varying
within 1$\sigma$ our results (eq.\ref{eq:res1}) is: $W\simeq 0.02$.
By adding $W$ to our estimated (raw) $w(\theta)$ and fitting again the
model correlation function we find: 
\be\label{eq:res2}
\theta_{\circ} \simeq 44 \pm 20 \;\; {\rm arcsec} \;\;\; \gamma 
\simeq 2 \pm 0.25\;,
\ee
consistent within the errors, with our uncorrected results
(eq. \ref{eq:res1}). If we 
fix $\gamma=1.8$ we obtain $\theta_{\circ} =28 \pm 9$
arcsec. Due to the small effect of the 
{\em integral constraint} correction we
will use in the rest of the paper our {\em raw} $w(\theta)$ estimates.

Our results show that hard X-ray sources are strongly 
correlated, even more 
than the soft ones (see Vikhlinin 
\& Forman 1995; Yang et al. 2003; Basilakos et al. in preparation). 
Our derived angular correlation length $\theta_{\circ}$ is in rough agreement,
although somewhat smaller (within 1$\sigma$) with the
{\em Chandra} result of $\theta_{\circ}=40 \pm 11$ arcsec
(Yang et al. 2003). 
The stronger angular clustering with respect to the soft
sources could be either due to the higher flux limit of the hard
XMM-{\em Newton} sample, resulting in the selection of relatively 
nearby sources, or could imply an association of our hard X-ray
sources with high-density peaks
%, and thus they are highly biased tracers 
of the underline matter distribution. 
To test the latter suggestion we have measured the 
cluster - hard X-ray sources cross-correlation
function ($w_{c,hard}$) using either the Goto et al. (2002)
clusters, detected in the multicolour optical SDSS data,
or the X-ray clusters that we have detected on our fields (Gaga et
al. in preparation).
%In figure 3 we present
%$w_{c,hard}(\theta)$ (solid points) and the 1-$\sigma$ 
%(continuous line) and 2-$\sigma$ (dashed line) confidence levels
%using the Goto et al (2002) clusters. 
We find no significant cross-correlation signal 
[$w_{c,hard}(\theta < 300^{''})\simeq  -0.15 \pm 0.19$], a
fact that weakens the suggestion of association of the hard X-ray
sources with high-density peaks. However, we must stress that the null result
could be artificial, due to small number statistics.

\subsection{The spatial correlation length using $w(\theta)$}
The angular correlation function $w(\theta)$ can be obtained
from the spatial one, $\xi(r)$, through the Limber transformation
(Peebles 1980). If the spatial correlation function is
modeled as
\begin{equation}
\xi(r,z)=(r/r_{\circ})^{-\gamma} (1+z)^{-(3+\epsilon)} \;,
\end{equation} 
then for a flat Universe the amplitude $\theta_{\circ}$ in two 
dimensions is related to the 
correlation length $r_{\circ}$ (see Efstathiou et al. 1991) in 
three dimensions through the equation: 
\begin{equation}
\theta_{\circ}^{\gamma-1}=H_{\gamma}r_{\circ}^{\gamma}
\left(\frac{H_\circ}{c}\right)^\gamma 
\int_{0}^{\infty} \left( \frac{1}{N}\frac{{\rm d}N}{{\rm d}z}
\right)^{2} \frac{E(z)}{x^{\gamma-1}(z)}  (1+z)^{-3-\epsilon+\gamma}
{\rm d}z \;,
\end{equation}
where $x(z)$ is the proper distance, 
$E(z)=\sqrt{\Omega_{\rm m}(1+z)^{3}+\Omega_{\Lambda}}$ is the element
of comoving distance and $H_{\gamma}=\Gamma(\frac{1}{2})
\Gamma(\frac{\gamma-1}{2})/\Gamma
(\frac{\gamma}{2})$.  
Note that if $\epsilon=\gamma-3$, the clustering is constant in comoving 
coordinates (comoving clustering) while if $\epsilon=-3$ the 
clustering is constant in physical coordinates.
We perform the above inversion in the framework of either the {\em
  concordance} $\Lambda$CDM cosmological model 
($\Omega_{\rm m}=1-\Omega_{\Lambda}=0.3$, $H_{\circ}=70$km s$^{-1}$ 
Mpc$^{-1}$) or the Einstein-de Sitter model.

The redshift distribution ${\rm d}N/{\rm d}z$ and the predicted 
total number, $N$, of the X-ray sources which enter in eq. 5
can be found using the hard band luminosity
functions of Ueda et al. (2003) and of Boyle et al. (1998).
We also use different models for the evolution of the hard
X-ray sources: a pure luminosity evolution (PLE) or the more realistic
luminosity dependent density evolution (LDDE; Ueda et al 2003). In
Figure 3 we show the expected redshift distributions of the
hard X-ray sources for three different luminosity functions and
evolution models. Both the Boyle et al (1998) and
Ueda et al (2003) luminosity functions with pure luminosity evolution
give relatively similar ${\rm d} N/{\rm d} z$ distributions. 
However, the LDDE model gives an ${\rm d} N/{\rm d} z$ 
distribution shifted to much larger redshifts with a 
median redshift of $\bar{z} \simeq 0.75$ (see also Table 1).
%The expresion $W(N,z)\equiv N^{-1} {\rm d}N/{\rm d}z$, 
%introduced in the integral equation of clustering is strongly 
%affected by the redshift selection function of our survey. 
%In Fig. 3 we present the 
%observed (histogram - having redshifts for a subsample of 34
%hard X-ray sources) and theoretical (using 
%the Ueda et al. 2003 luminosity function) $W(N,z)$ distributions.
%The observed $W(N,z)$ distribution shows a $z>1$ tail and is not
%greatly different from the predicted one. 

For the comoving clustering model ($\epsilon=\gamma-3$) 
and using the LDDE evolution model, we estimate the hard X-ray source
correlation length to be: $r_{\circ}=19 \pm 3 \; h^{-1}$ Mpc
and $r_{\circ}=13.5 \pm 3 \; h^{-1}$ Mpc 
for $\gamma=1.8$ and $\gamma=2.2$ respectively.
While if $\epsilon=-3$ the corresponding values
are: $r_{\circ}=11.5 \pm 2 \; h^{-1}$ Mpc and 
$r_{\circ}=6 \pm 1.5 \; h^{-1}$ Mpc, respectively.
%Note that there is a further contribution to the uncertainty of $r_{\circ}$ 
%from the uncertainty of the LDDE luminosity function parameters,
%estimated to be $\Delta_{\rm XLF} \simeq 1 \; h^{-1}$\,Mpc. 
In Table 1, we present the values of the correlation
length, $r_{\circ}$, resulting from Limber's inversion for different
luminosity function and evolution models.

These estimated clustering lengths (for $\gamma=1.8$)
are a factor of $\gtrsim 2$ larger than the 
corresponding values of the Lyman break galaxies (Adelberger
2000), the {\em 2dF} (Hawkins et al. 2003)
and SDSS (Budavari et al. 2003) galaxy distributions as well as
the {\em 2QZ} QSO's (Croom et al. 2001). However, 
the most luminous, and thus nearer, {\em 2QZ} sub-sample ($18.25<b_{j}<19.80$) 
has a larger correlation length ($\sim 8.5 \pm 1.7 \; h^{-1}$ Mpc) than the
overall sample (Croom et al. 2002), in marginal agreement 
with our $\epsilon=-3$ clustering evolution results.

The large spatial clustering length of our hard X-ray sources 
can be compared with that of Extremely Red Objects (EROs) and luminous
radio  sources (Roche, Dunlop \& Almaini 2003; Overzier et 
al. 2003; R\"{o}ttgering et al. 2003) which are found to be 
in the range $r_{\circ} \simeq 12 - 15 \; h^{-1}$ Mpc. 
The possible association of EROs with high-$z$ massive ellipticals and
of luminous radio sources with protoclusters (for a review see
R\"{o}ttgering et al. 2003 and references therein) suggests that our
hard X-ray sources could trace the high peaks of the underline mass
distribution (see also Yang et al. 2003). 
%This is also in agreement
%with the results of  {\em Chandra} (Yang et al. 2003) that also suggest
%an association of hard X-ray sources with high density regions. 

\section{Conclusions}
In this paper we explore the  clustering properties of hard (2-8\,keV)  
X-ray selected sources using a wide area ($\rm \approx 2\,deg^2$)
shallow [$f_X(\rm 2-8\,keV)\approx 10^{-14}\rm \, erg \, cm^{-2} \, 
s^{-1}$] XMM-{\it Newton} survey. Using an angular correlation function
analysis we measure a clustering signal at the $\sim 4\sigma$ confidence
level. Modeling the  angular correlation function by a power-law, 
$w(\theta)=(\theta_{\circ}/\theta)^{\gamma-1}$, we estimate  
$\theta_{\circ}=48.9^{+15.8}_{-24.5}$\,arcsec and  $\gamma=2.2\pm
{0.30}$. Fixing the correlation function 
slope to $\gamma=1.8$ we estimate
$\theta_{\circ}=22.2^{+9.4}_{-8.6}$\,arcsec. Using 
a variety of luminosity functions and evolutionary models
the Limber's inversion provides correlation lengths which are in the
range $r_{\circ} \sim 10 -19 \; h^{-1}$ Mpc, typically larger than
those of galaxies and optically selected QSO's but
similar to those of strongly clustered populations, like EROs and
luminous radio sources. 
%A cross-correlation function analysis
%of our X-ray sources with optical and X-ray clusters, found in the 
%area of our survey did not provide a positive and significant signal.

\acknowledgments
We thank the anonymous referee for useful suggestions.
This work is jointly funded by the European Union
and the Greek Government in the framework of the program 'Promotion
of Excellence in Technological Development and Research', project
{\em 'X-ray Astrophysics with ESA's mission XMM'}. Furthermore, MP
acknowledges support by the Mexican Government grant No
CONACyT-2002-C01-39679.

\newpage

\begin{table}[h]
\caption[]{The hard X-ray sources correlation length ($r_{\circ}$ in $h^{-1}$ Mpc)
 for different pairs of ($\gamma, \epsilon$) and 
for the different luminosity functions and evolution models. The last
column indicates the predicted median redshift, from the specific luminosity function
 used. The bold letters deliniate the prefered cosmological model and
 the most updated luminosity function.}
\vspace{1cm}

\tabcolsep 3pt
\begin{tabular}{cccccccc} \hline
LF & Evol. Model & ($\Omega_{m},\Omega_{\Lambda}$) &
$r_{\circ}$ $(1.8,-1.2)$&
$r_{\circ}$ $(1.8,-3)$
& $r_{\circ}$ $(2.2,-0.8)$  
& $r_{\circ}$ $(2.2,-3)$ & $\bar{z}$ \\ \hline
Boyle & No evol. & (1,0)   & 11.5$\pm 2.00$    & 9.0$\pm 1.5$ & 7.3$\pm 1.2$  & 6.0$\pm 1.5$&0.45 \\
Ueda  & No evol. & (1,0)   & 9.5$\pm 1.5$ & 7.5$\pm 1.0$& 6.5$\pm 1.5$  & 5.0$\pm 1$ &0.40 \\
Boyle & PLE      & (1,0)   & $13.0 \pm 3.0$   & 10.0 $\pm 2.0$& 8.0$\pm 2.0$  & 6.8 $\pm 1.5$ &0.50 \\
{\bf Ueda}  & {\bf PLE} & {\bf (0.3,0.7)} & ${\bf 13.0 \pm 2.0}$ &
${\bf 9.3\pm 1.6}$ & ${\bf 9.0\pm 2.0}$ & ${\bf 6.7\pm 1.5}$ & {\bf 0.45} \\
{\bf Ueda} & {\bf LDDE} & {\bf (0.3-0.7)} & ${\bf 19.0 \pm 3.0}$ & ${\bf
  11.5 \pm 2.0}$& ${\bf 13.5\pm 3}$ & ${\bf 8.5 \pm 2}$&{\bf 0.75} \\
Ueda  & LDDE     & (1,0)   & $15.0 \pm 2.5$ & 7.8 $\pm 1.5$& 11.0$\pm 2.5$ & 6.0$\pm 1.5$&0.80 \\ \hline
\end{tabular}
\end{table}

\newpage

\begin{inlinefigure}
\label{fig_lognlogs}
\epsscale{0.8}
\plotone{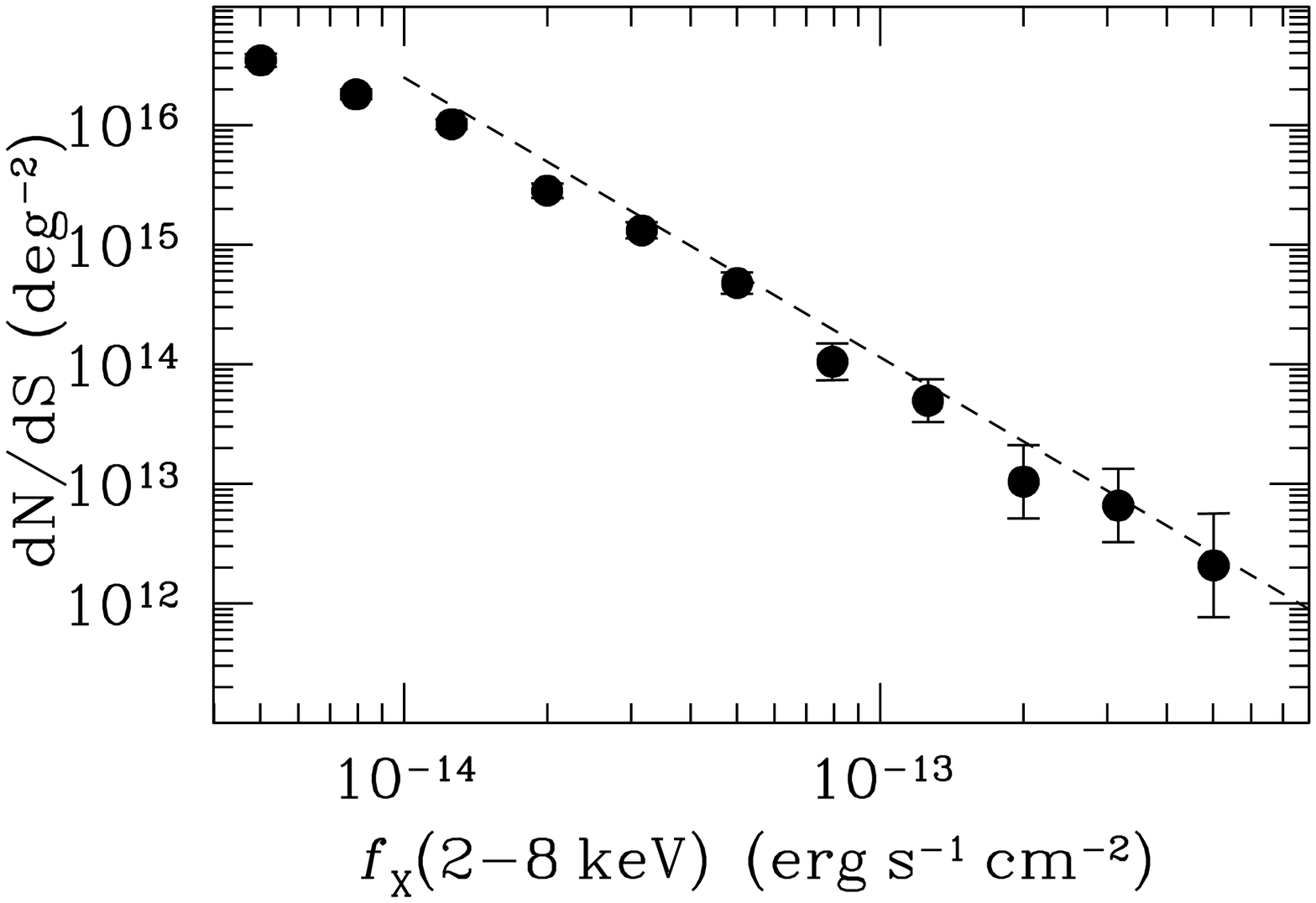}
\figcaption{The hard band (2-8\,keV) differential number counts from the
present survey in comparison with those of Baldi et al. (2002; dashed line).}
\end{inlinefigure}

\begin{inlinefigure}
\label{fig_w}
\epsscale{0.8}
\plotone{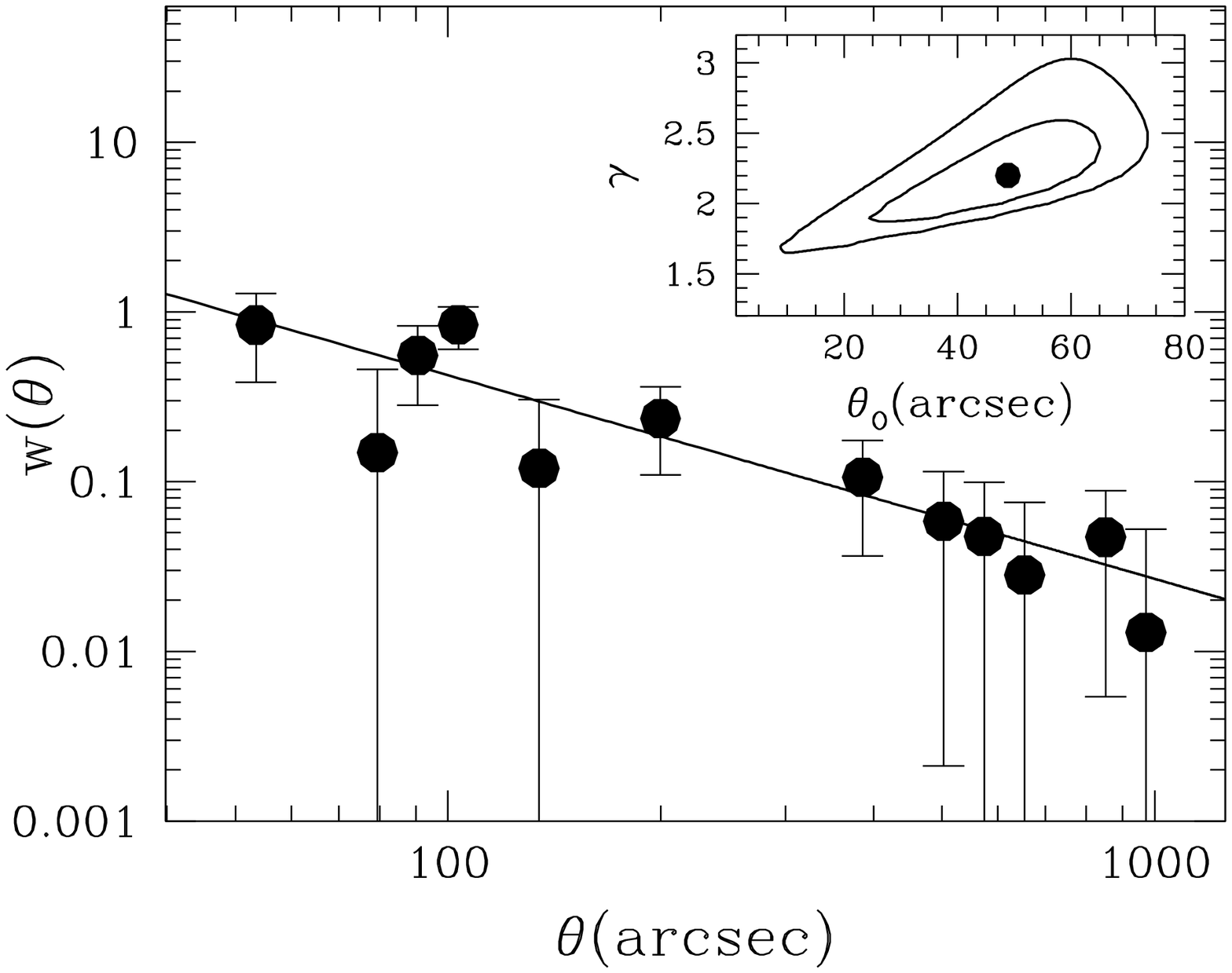}
\figcaption{The two-point angular correlation function 
of the hard (2-8 keV) X-ray sources. The line
represents the best-fit power law 
$w(\theta)=(\theta_{\circ}/\theta)^{\gamma-1}$ with 
$\theta_{\circ} = 48.9$ arcsec and $\gamma=2.2$. Insert: 
Iso-$\Delta \chi^{2}$ contours in the $\gamma$-$\theta_{\circ}$ 
parameter space.}
\end{inlinefigure}

%\begin{inlinefigure}
%\epsscale{0.8}
%\plotone{cross_goto.ps}
%\figcaption{The two-point cross-crorrelation function between
%Goto et al (2002) clusters and our hard X-ray sources. The deliniated
%region indicates the one and two $\sigma$ levels, estimated from a
%large number of Monte-Carlo simulations of a 
%random distribution of ``cluster'' centers, having the same number as
%the Goto et al (2002) clusters, and an off-axis selection function similar
%to that of our X-ray sources.}
%\end{inlinefigure}

\begin{inlinefigure}
\label{fig_nz}
\epsscale{1.1}
\plotone{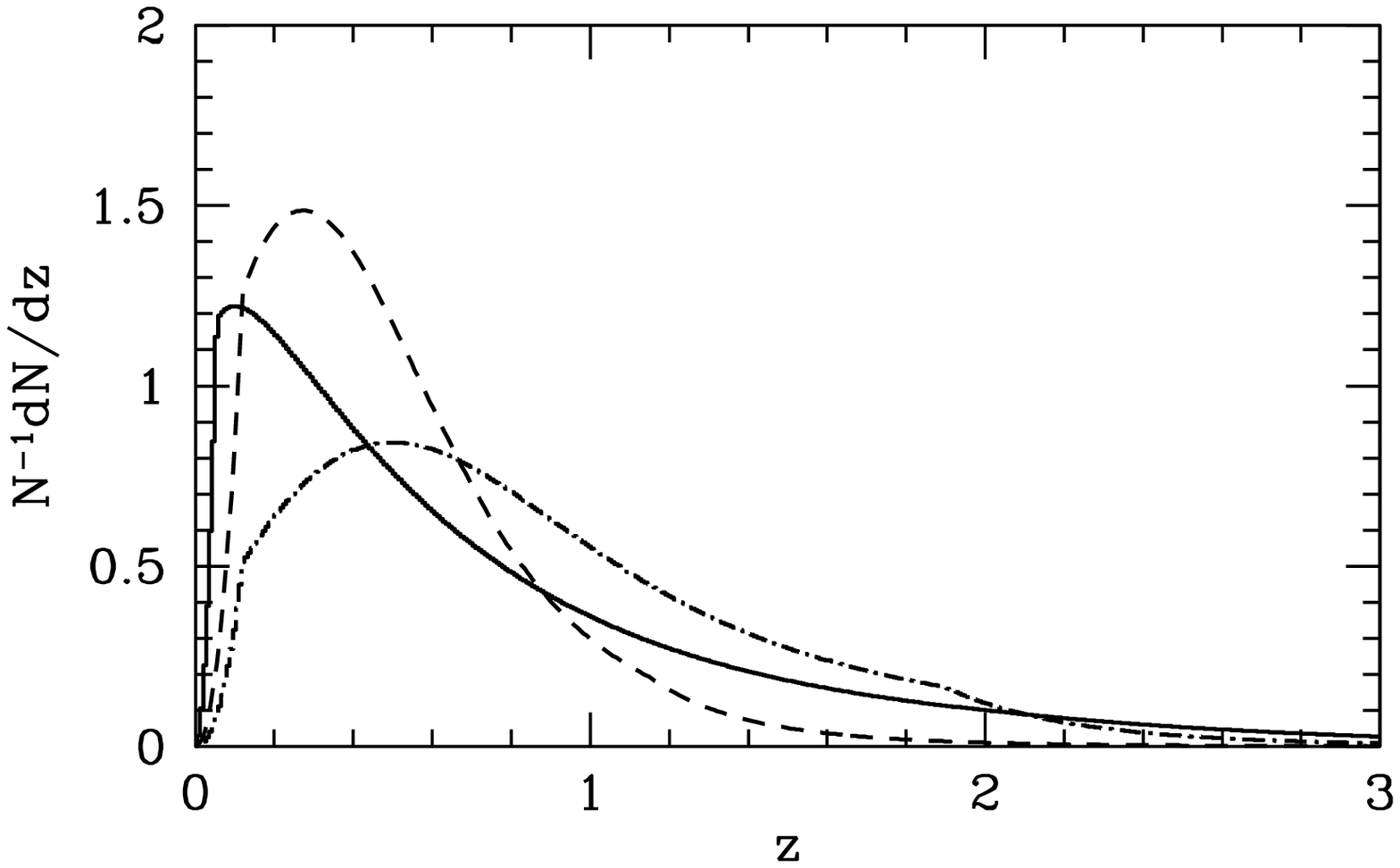}
\figcaption{The predicted $N^{-1}{\rm d}N/{\rm d}z$ distribution for
  three different LF models: (a) Boyle et al. (1998) with PLE
  (continuous line), (b) Ueda et al (2003) with PLE (dashed line), (c)
  Ueda et al (2003) with LDDE (dot-dashed line).}
\end{inlinefigure}

\end{document}